\documentclass[runningheads]{llncs}
\usepackage[T1]{fontenc}
\usepackage{hyperref}
\usepackage{graphicx}
\usepackage{physics}
\usepackage{yquant}
\usepackage{todonotes}
\usepackage{subcaption}
\begin{document}
\title{Explainable Quantum Machine Learning for Multispectral Images
 Segmentation:  Case Study}
 \titlerunning{XQML for MSI Segmentation: Case Study}
%
%
\author{
	Tomasz Rybotycki$^{1,2,3[0000-0003-2493-0459]}$ \and
	Manish K. Gupta $^{2[0000-0003-4226-0364]}$ \and
	Piotr Gawron$^{1,2[0000-0001-7476-9160]}$
}
\authorrunning{Tomasz Rybotycki, Manish K. Gupta,  Piotr Gawron}
%

\institute{
	Systems Research Institute Polish Academy of Sciences, ul.~Newelska 6, 01-447 Warsaw, Poland
	\and
	Nicolaus Copernicus Astronomical Center, Polish Academy of Sciences, ul. Bartycka 18, 00-716 Warsaw, Poland
	\and
	Center of Excellence in Artificial Intelligence, AGH University, al.~Mickiewicza 30, 30-059 Cracow, Poland
}
\maketitle              
\begin{abstract}

The emergence of Big Data changed how we approach information systems
engineering. Nowadays, when we can use remote sensing techniques for Big
Data acquisition, the issues such data introduce are as important as ever. One
of those concerns is the processing of the data. Classical methods often fail to
address that problem or are incapable of processing the data in a reasonable
time. With that in mind information system engineers are required to investigate
different approaches to the data processing. The recent advancements in noisy
intermediate-scale quantum (NISQ) devices implementation allow us to investigate their application to real-life computational problem.  This field of study is called quantum (information) systems engineering and usually focuses on technical problems with the contemporary devices. However, hardware challenges are not the only ones that hinder our quantum computation capabilities. Software limitations are the other, less explored side of this medal. Using multispectral image segmentation as a task example, we investigated how difficult it is to run a hybrid quantum-classical model on a real, publicly available quantum device. To quantify how and explain why the performance of our model changed when ran on a real device, we propose new explainability metrics. These metrics introduce new meaning to the explainable quantum machine learning; the explanation of the performance issue comes from the quantum device behavior. We also analyzed the expected  money costs of running similar experiment on contemporary quantum devices using standard market prices.


\end{abstract}

\section{Introduction}

Interdisciplinary character of information systems engineering (ISE) is an well-known fact \cite{ISE2005}. It is closely related to software engineering (SE) \cite{IS-SE} and information technology (IT) \cite{Navigating}. While relation of ISE to both those disciplines is relevant in the context of this work, the connection to IT highlights an interesting shift that currently occurs in the field. IT became a prominent part of ISE due to relentless digitization of essentially every field \cite{Digitalization,Navigating}. This process is so extreme that it made technology the core around which the organizational processes and structures are built \cite{DigitalizationImpact,Navigating}. If we additionally consider the challenges introduced by industry 4.0 \cite{Industry4} (big data, processes automation) and the advent of industry 5.0 \cite{Industry4v5,Industry5XAI} (human-centric approach, XAI), we can clearly see how a dire need for ISE methodology re-evaluation \cite{Navigating} arise. 
	
There's a consensus among scientists from different fields that the solutions to all ISE challenges cannot be purely technical. Ethicists argue that some of the problems, such as so-called responsibility gap, require human participation --- at least in some limited fashion \cite{ResponsibilityGap}. Socio-technical approaches are more common, especially in the context of IS and ISE \cite{SociotechnicalISE,SociotechnicalISE2}. However, if we consider the current ISE research agenda \cite{Navigating} we may notice that some of its themes --- (4) the impact of emerging technologies on ISE and (5) the effects of artificial intelligence on ISE, in particular --- leave some room for technical solutions \cite{QSE,RPA,Blockchain,ISEAI,XAIISE}. The focus of this work is on the technical solution to the Big Data problem --- quantum technologies.

In the context of ISE methodology re-evaluation, quantum systems engineering (QSE) \cite{QSE} provides a fresh view. Its methodology differs from that of ISE or even system engineering \cite{QSEPhD}. The reason is fundamental --- it spurs from the quantum properties of such systems \cite{QSEPhD}. While QSE technically refers to all quantum technologies, those encompass quantum computing (QC) \cite{QSEQC,QCISE} or even literally Quantum Information Systems (QIS) \cite{QIS}. QIS are important subset of IS, not only because they are expected to outperform classical IS for some computational tasks (so-called quantum advantage) \cite{QA}, but also because they can naturally handle quantum data \cite{QData}. An excellent example of such systems are satellites for quantum key distribution (QKD) \cite{QKD1,QKD2,QKD3} --- a quantum cryptography protocol that requires communication via quantum channels \cite{QKD}. Quantum machine learning (QML) techniques are also widely used in the earth observation research \cite{QHSI,QHSI2}, although those approaches are rarely tested on actual quantum devices (compare simulation vs. hardware experiments in \cite{QMLUse,QMLUse2}). QML is promising, but there's still a lot to be done.

Moving the frontier of QML requires us to start
performing the experiments on the real quantum devices. Only this way
can we design test noise robustness of current QML algorithms and
fully use the power of noisy intermediate-scale quantum (NISQ) devices. In this work, we present a case study
of using publicly available gate-based quantum computers for multispectral
satellite image segmentation. The case study will encompass the whole experimental pipeline, from noiseless simulations to inference of the actual quantum hardware. Our focus during the study will primarily be on QML software. Our goal is to reproduce typical workflow of a QML researcher and report all difficulties that during the experiment. During the research, we will 
interact (although indirectly in some cases) with essentially every layer of the standard QC architectural framework \cite{QIS}.

QML model training on a real machine is practically impossible due to
the resource limitations. Access to quantum machines are costly and restricted. They can, however, be used for inference. In
that case there's only one way to proceed --- use classical computer
to simulate the training. The goal would be to simulate the training environment in such way that it would be as close to real device as possible. The simulator should consider the target device topology, its noise model and other available parametrization. We will call this approach a device-oriented training.

While the concept of device-oriented training sounds fairly simple, its
implementation requires --- among others ---  a successful communication
between different software libraries, each responsible for a specific 
task related to the training. During our experiments, we've noticed
that this requirement isn't as straightforward as one would assume. 
The problem is so prominent, that we had to resign from using one
of the quantum computers available to us, due to the libraries and/or
library versions incompatibility issues. These kinds of problems are
very important, although not much attention is given to them in the
literature. In \cite{QMLSoft} the authors compare QML libraries, but neither go into details nor report any issues. The finding of \cite{QSDev} shows that community is aware of some quantum software shortcomings, but again, without discussing the details of actual software errors.

Our investigations also shown that there's a certain gap, not only in the QML software, but also in the general discussion. We expected that the quality of our final model will be lower, when we run it on the real hardware. This is due to the noise present on the device. There are, however, no explainability metrics that would allow to rigorously highlight this issue.  Explainable artificial intelligence (XAI) techniques, that's been very popular in the context of industry 5.0 \cite{Industry5XAI} and ISE \cite{XAIISE}, are also used in QML \cite{XAIQMLComparison}. However, the only publications we found on that topic concerned application of classical XAI techniques to QML models \cite{XAIQML1,XAIQML2,XAIQML3}. Simulator-device results discrepancies were never discussed in the context of XAI for QML. 

With this paper we make several contributions. First and foremost, we present
a case study of complete QML hardware experiment pipeline (i. e. from noiseless simulators to real hardware) with a focus on the software-related issues. While the problems are or might be solvable, we claim that they shouldn't be present at all, thus their occurrence allow us to make a limited statement about the state of current QML software. Second of all, we will fill the gap in the explainable QAI, by introducing simple explainability metrics for analyzing the QML models behavior when ran on a real quantum device. We implement those metrics, thus also addressing the gap in the QML software. Moreover, we propose a notion of device-dependent training, which aims to limit the models performance drop, when the model is ran using targeted hardware. Finally we also estimate the money costs of running our experiment if the access to QPU were to be bought at market price. Concluding all our findings, we make a claim about the usability of QML models ran on contemporary quantum hardware.

This paper is organized as follows. In the following section, we discuss the example machine learning task we selected for the experiment. We also introduce the dataset we will use.  In section \ref{sec:approach} we will
describe our method. We introduce the hybrid quantum-classical model
that we've used and its different variations. Our focus will be particularly
on the concept of the device-oriented training. We also briefly summarize
our initial results in the context of the on-going research on the topic.
In section \ref{sec:metric} we introduce the explainability metrics we will
use to discuss the results obtained using the selected quantum device.
In the next section, we will discuss the experiments we've done and estimate their money cost. We will 
explain the setup of the experiment and  mention all the software-related issues we've encountered. Then, we will assess the performance of our model ran on both devices. We will investigate and explain the difference using the explainability metrics introduced in section \ref{sec:metric}. Finally, we
will conclude the paper with the interpolated insights about the status of contemporary QML software.

\section{Machine Learning task formulation}
\label{sec:data}

Considering the popularity of QML techniques in satellite (multi-) hyperspectral image analysis \cite{QHSI,QMLHSI,QMLRS}, we decided to select similar task as an machine learning (ML) problem example. In the context of satellite images analysis, image segmentation is of particular interest \cite{Daudt2018}. There are two main approaches to the image segmentation. First is a pixel-level one, when each pixel is analyzed separately and a patch classification, where larger groups of neighboring pixels are considered simultaneously
\cite{Cong2019,Seunghyeok2020}. Since one of our goal is to assess the quality of a QML model run on a quantum device, we wanted the task to be as simple as possible. Considering the quantum hardware limitations and potential data encoding issues, we decided to use the simpler, pixel-level approach.

\begin{figure*}
	\centering
	\includegraphics[scale=0.24]{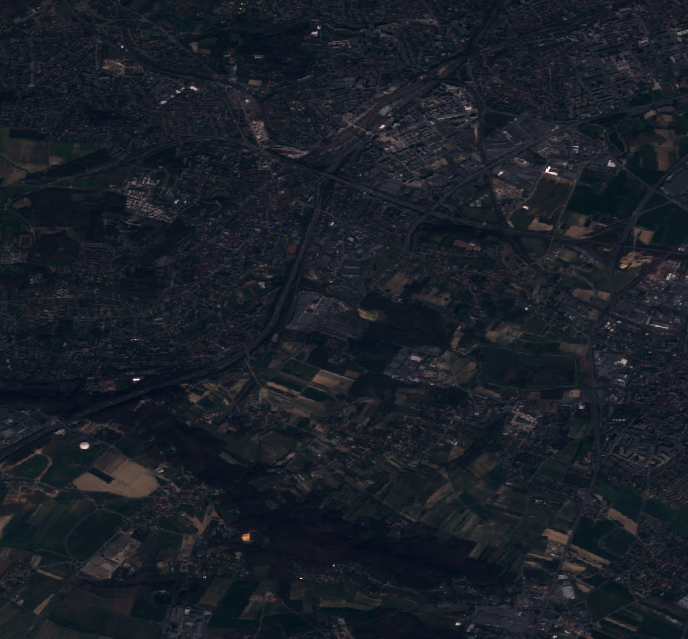}
	\includegraphics[scale=0.24]{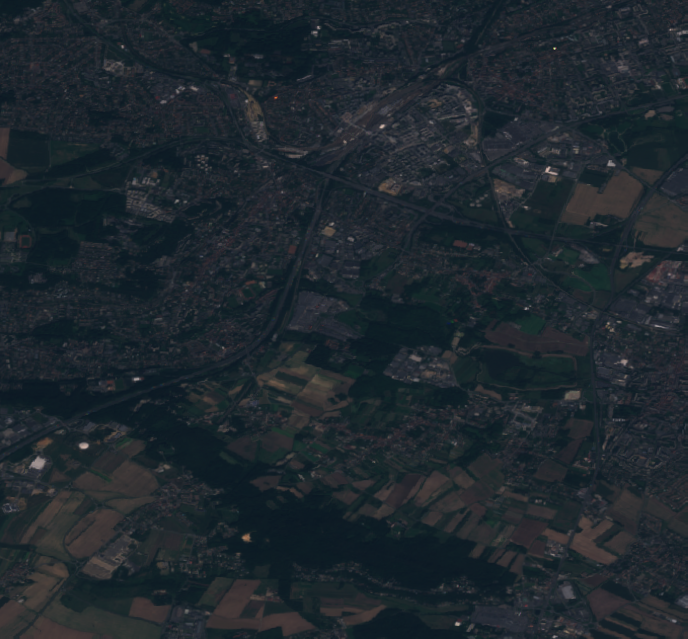}
	\caption{A pair of images from the ONERA dataset. They show Saclay, a city in France. The images were taken on 15 III 2016 (Left) and 29 VIII 2017 (Right).}
	\label{fig:ONERA}
\end{figure*}

For our study, we selected the Onera Satellite Change Detection Dataset \cite{ONERA}. This dataset consists of 24 pairs of multispectral images taken from the Sentinel-2 satellites, each image having 13 spectral bands and vary in spatial resolution from 10m to 60m. Each pair consists of the images presenting the same area, but in a different time (usually over a year later). The images were taken all over the world. The ML task for which ONERA dataset was prepared is change detection, i.e. given a pair of images find those regions on the image that are different in some predefined way. In the case of ONERA, the ground truths were prepared for urban changes detection, such as new buildings or new roads. An example of image pair from the ONERA dataset is presented in Fig. \ref{fig:ONERA}.

It is a well-known fact that change detection can be interpreted as both (semantic) segmentation \cite{CDtoSS} and binary classification task\cite{BinaryChangeDetection}. The interpretation of the results in the latter formulation is straightforward --- for a given pixel there either was a change or not. Since binary classification is considered one of the simplest ML tasks \cite{binclas}, we used this formulation when were preparing out model.

\section{Proposed approach}
\label{sec:approach} 

One of the main goals of this study was to investigate the potential issues that arise when one wants to run a QML model on a quantum hardware. However QML models are a very broad class of ML models. In particular, they also include hybrid quantum-classical models, i.e. models where quantum component is only their part. Several QML libraries implement functions that allow one to transform their variational quantum circuit (VQC), known as quantum neural network (QNN), into programming objects compatible with classical ML models (e.g. \texttt{TorchLayer}s of PyTorch \cite{Imambi2021}). We decided that this additional layer would be both interesting to investigate and would add more expressiveness to our model, so we decided to use a hybrid quantum-classical model in our research. For the sake of brevity, we will tend to use quantum model instead of hybrid quantum-classical.

\begin{figure*}
	\centering
	\includegraphics[scale=0.5]{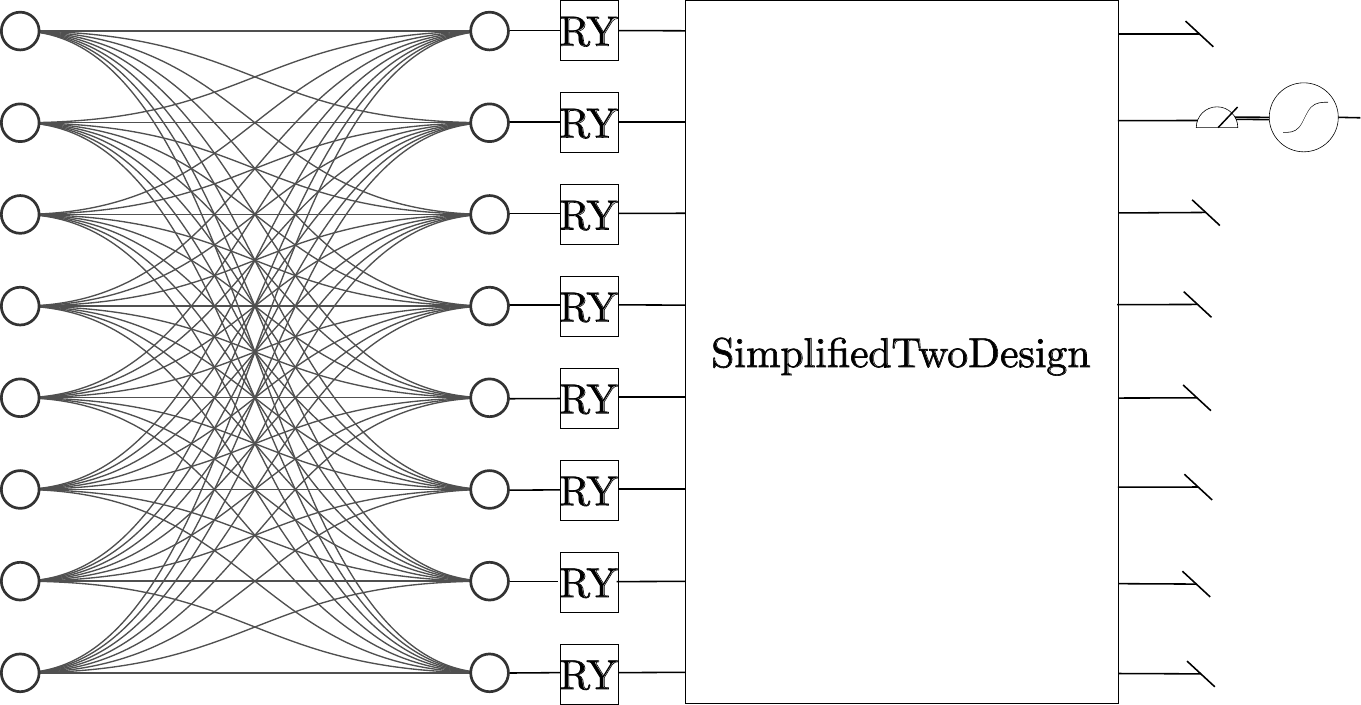}
	\caption{A schematic of the model we used in our pixel-level
		change detection experiments. The input date first passes
		through a single layer of a classical perceptron. It is then encoded
		into a quantum circuit via a layer of $RY$ gates. Then it is processed
		by three \texttt{SimplifiedTwoDesign} layers.
		We measure the expectation value of a selected
		qubit, pass it through a sigmoid function and return the final value
		as the output of the model. See appendix \ref{app:layers} for the quantum layers details.}
	\label{fig:model}
\end{figure*}

The structure of our final model is presented in Fig. \ref{fig:model}. It consists of a classical single layer perceptron, followed by a QNN layer (on Fig. \ref{fig:model} there one can see specific components of this layer), dychotomic single qubit measurement in the computational ($\{\ket{0}, \ket{1}\}$) basis and sigmoid activation function. The results of the latter is then rounded (to either 0 or 1) and considered the classes returned by the model.

To find a quantum layer with the best performance, we tried different configurations of standard embedding methods and QNN layers \cite{PennyLaneDocumentation},
along with the Bellman circuit layers \cite{Sebastianelli} that were previously used for a similar task. In Table \ref{tab:acc}, we present accuracy values of 12 models that we tested. Those values were obtained using noiseless quantum device simulators. That's because such experiments would be extremely costly (both money and time-wise) on the real hardware. This was also the early stage of the experiments, so simulator was more reasonable. Notice that amplitude embedding and SimplifiedTwoDesign layers worked best (both separately and together). Therefore, the final QNN layer consisted of an amplitude embedding layer, followed by three \texttt{SimplifiedTwoDesign} layers. We detailed the quantum circuit representations of each layer in appendix \ref{app:layers}. The model implementation is available on GitHub \cite{Code}.


\begin{table}
	\caption{Overall accuracy of the model with respective quantum
		layers. Rows contain information about the embedding we used, whereas
		the columns describe the circuit's layers. SEL and STD denote Strongly
		Entangling Layer and Simplified Two Design respectively. Best accuracy
		scores are bolded.}
	\label{tab:acc}
	\centering
	\begin{tabular}{| c || c | c | c | c |} 
		\hline
		& \textbf{Bellman} $\times 4$& \textbf{SEL} $\times 1$& \textbf{SEL} $\times 3$ & \textbf{STD} $\times 3$\\ 
		\hline \hline 
		\textbf{Amplitude} & \textbf{0.74} & 0.73 & 0.73 & \textbf{0.74} \\ 
		\hline 
		\textbf{Angle (X)} & 0.52 & 0.66 & 0.73 & 0.73 \\
		\hline
		\textbf{Angle (Y)} & 0.52 & 0.67 & 0.71 & \textbf{0.74} \\
		\hline
	\end{tabular}
	
\end{table}

\bgroup
\def\arraystretch{1.5}
\setlength\tabcolsep{5pt}
\begin{table}[ht!]
	\caption{Metadata of the QNN layer of our model transpiled to a specific device (\texttt{ibm\_brisbane}).}
	\label{tab:meta}
	\centering
	\begin{tabular}{| c | c | c | c | c | c | c | c |} 
		\hline
		\textbf{Embedding} & \textbf{QNN Layers} & \textbf{\# qubits} &  \textbf{depth} & \textbf{\# rz} & \textbf{\# sx} & \textbf{\# ecr} & \textbf{\# x} \\
		\hline
		Amplitude & 3 x STD & 8 & 44 & 143 & 82 & 21 & 6  \\
		\hline
	\end{tabular}
\end{table}
\egroup


When we compare the results of our model
with the baseline classical model from \cite{daudt2018urban},
we find that quantum model performs poorly in terms of classification
accuracy. We'd like to point out that we didn't run the tests for the
classical model --- we took the results directly from
\cite{daudt2018urban}. The overall accuracy of out pixel-level model is
about 15 percent points worse than that of the classical one. This is quite easy
to explain, considering that our model had less than 200
parameters, compared to a million in the classical model.  To support this claim
we conducted additional experiment. We reduced the classical
U-Net model \cite{Daudt2018} to a single layer of encoder and decoder. This
dropped the number of trainable parameters to 1312. For a patch size of
$2\times 2$ pixels, the reduced model accuracy on test dataset per class was
97.9\%  (change) and 25.5\% (no change). The drop is significant, however,
we could not use these results for a fair comparison, because the number of
parameters in the classical network is still over 6 times higher.

To ensure the results obtained when running the model using quantum
hardware are as good as possible, we wanted to train our selected model
in an environment as close to the target one, as possible. This required
us to select a target quantum computer, and consider its topology,
native gates and noise model during the training. In that sense, we 
consider this approach a device-oriented training.

We decided to divide the training into three different phases. We start
with the simplest simulation, assuming interconnection between every pair
of qubits and no noise. In second phase, we add the information about
the device topology and the device-native gates. It is important to
consider topology before the noise, because different physical qubits
have different noise parameters. The noise is also typically gate-dependent
and the process of circuit transpilation may significantly influence the
circuit architecture. Also, noise model is usually straightforward to apply
once virtual-physical qubits mapping is known and the circuit is transpiled.

For our hardware experiments, we knew we would have access to quantum
computers from two different providers --- IBM and IQM. Both of them use
Qiskit \cite{qiskit2024} to communicate with the target device. Our main
library --- PennyLane \cite{PennyLaneDocumentation} --- can handle
Qiskit, thus the transition between the hardware should be seamless. However,
due to variety of issues, starting with resource access, through quantum
gates implementations status between different versions of the libraries and
finally due to different observable encoding strategies, we could not run our
model on the IQM device simulator. We tried using qiskit-machine-learning
library to bypass the PennyLane-Qiskit-IQM configuration issues, but it
didn't help. In the end, we were only able to consider the topology, but the
native gates and noise model was inaccessible. Thus, we were forced to run
our experiments on the IBM device only. 

Although our device-oriented training was smoother for IBM device,
we were not able to complete it. Although the toy models we tested
could run forward- and backward passes using a noisy simulator, they
were not able to complete a single epoch of the training. The common
issue here were the out-of-memory problems, even on our computational
clusters. While our model wasn't overly complicated, neither deep nor 
with a large number of qubits, noisy simulation deemed too much for
current software. This highlights the limitations of current QML software
and proves that the QML software requires more attention.

\section{Noisy Hardware Explainability Metrics}
\label{sec:metric}

As we mentioned in the introduction, there's a gap in XAI techniques for QML. There currently are no XAI methods that would allow one to explain the change in models performance when comparing models ran using simulator with ones ran on real hardware. Since current devices are noisy, we obviously expect the results to worsen. However, we still need metrics that would allow us to investigate and ensure that's indeed the case. 

The need for such metrics can be easily explained. Imagine you've got a terrible binary classifier and a balanced dataset. The accuracy of you classifier on noiseless simulator is 30\%. But what would happen if you were to run it on a real device? Would the accuracy drop, as one may naively expect, or would it increase? The answer to that question might not be obvious, as we will see in the experimental section \ref{sec:exp}. To fill the gap in XAI methods for QML and provide the tools to answer such questions, we propose three metrics: Sureness, Confidence and Imbalance. 

Let's start with sureness. It describes how sure the model
was about the class assignment. Its value is bound to [0, 1]. The closer the model output was to
0.5 (which is essentially the demarcation line between the classes), the less sure about the assignment it was. Formally we denote model sureness as:

\[ \mathrm{Sureness}(M) = 2 \cdot \sum_{s \in S} \frac{|M(s) - 0.5|}{|S|}, \] 

\noindent where $S$ denotes the samples set, $M$ is the model for
which Sureness is computed, and $|\cdot|$ denotes either absolute
value (for scalars) or power of the set. The factor 2 is used to set the
sureness values range to [0, 1]. 

Another explainability metric that we propose is Confidence. It is a measure that aggregates information from Sureness and accuracy metrics. Intuitively, it describes how confident we can be in the model prediction. We assess not only if the model was correct or wrong, but also how confident it was about it. Formally
we can denote the confidence as follows:

$$
\mathrm{Confidence}(M, L, S) = 1 - \sum^N_{ i = 1 } \frac{|l_i - M(s_i)|}{N},
$$

\noindent where $N := |S|$, $L$ is the set of labels with values either 0
or 1 and the rest of the symbols is the same as in Sureness.  We can see that the confidence is 0, when the model is always wrong and maximally sure
of its decision, and 1 when it is maximally sure, but always rightfully so.

Finally we propose class prediction imbalance (or simply imbalance) as the last explainability metric for QML. Although in its current form it is only applicable for binary classification, we believe that similar metrics could be proposed for the other tasks. They are, however, out of scope of this paper. Imbalance is an integer value from the $[-\infty, \infty]$ interval, that is the difference of the
correct predictions of class 0 and class 1 by the model. With the information about the shots number $n$, the imbalance limits can be bounded to $[-n, n]$.
Formally we can denote it as

\[
\mathrm{Imbalance}(M, L, S) = n_0(M, L, S) - n_1(M, L, S),
\]

\noindent where $n_i$ is the number of samples from $S$ with the label $i$
classified correctly by the model $M$.  The greater the
imbalance's absolute value, the greater was model tendency to correctly
assign only one of the labels. The sign of the imbalance shows which
class was classified more (less) often.

To conclude this chapter, let us point out that the explainability metrics we propose could still be used alongside the standard XAI metrics. In this way, the metrics we propose constitute an extension to the XAI techniques for QML models. They allow one to broaden the explainability scope to also include hardware failures.

\section{Experiments}
\label{sec:exp}

In our research, we used standard QML libraries --- PennyLane and Qiskit ---
with their plugins. During the study we encountered a number of
software-related issues, i.e.: libraries version dependencies conflicts,
lack of support for device-native gates / standard layers, data batching issues,
libraries integration issues, remote resources / access handling
issues and finally out-of-memory errors that halted our pursue for complete
device-oriented training. Many of the issues still remain unaddressed. 

The software issues we encountered, limited what quantum devices we could use. Despite having access to both IBM and IQM devices, we were not able to run our model on IQM devices due to incompatible observable coding scheme between IQM and Pennelane. With only IBM devices to select from, we decided to use 
(\texttt{ibm\_brisbane}) for our experiments.  Considering only the target device  topology and native quantum gates set, we managed to automatically reconfigure and train our model. In the end, we were able to run the model quantum hardware.
 
While analyzing the results, we noticed a curious behavior of our
model. Contrary to our expectations, the classification
accuracy obtained using real device was much better than
the accuracy of simulated model, here: 43\% and 33\% respectively. 
However, the latter accuracy was also significantly lower than expected
after the training (72\%, see table \ref{tab:acc}). We investigated this issue
further and found out the reason. Running simulator-trained model on
a real device requires rebuilding of the quantum layer with the 
\texttt{PennyLane} device set to the target remote device. While we
ensured that the model is properly saved and loaded, and that the
weights of each layer of the model are set prior to the experiment, 
we wrongly assumed that the weights we set are automatically
applied to the model during forward pass. That is not the case. 
The weights of \texttt{qml.TorchLayer} are set during creation of this 
\texttt{TorchLayer}, and unless explicitly specified, they will be set 
to random weights. Ultimately, the model we ran on the real device, 
was the one with untrained quantum layer. We already fixed this problem
in our scripts.

Our initial thought was to rerun the experiments. Due to the time limitations, enforced on us by the project, that proved impossible. We therefore decided to investigate whether the results we've got can be used to gain any qualitative insights about the results. The answer was positive.

We started with the money cost estimation of the experiment. The goal here was twofold. First, we believe that such information would be of interests to the parties interested in using quantum devices for similar tasks. Second, we wanted to have this information at hand in case we had to repeat the experiments after the project deadline. The market prices for the IBM are defined in the per minute basis, so we started with estimating the QPU usage time. We gathered the date over 5 days --- 18 I 2025 -- 23 I 2025. The total QPU usage in that time was over 14 hours\footnote{Precisely 14 hours and 11 minutes.}. During that time we managed to classify only 1815 samples (918 and 897 of class 0 and 1 respectively) using 1024 real device shots for each sample. Average computation time per sample was 2 minutes and 13 seconds. Total money cost\footnote{96 USD per minute, as of Pay-as-you-go plan.} of obtaining those results using
IBM Quantum machines would be 81696 USD. Full test dataset consists of 31256 samples, which means that we could only process roughly 6\% (5.806 \%) of the dataset. Estimated cost of full test dataset processing would be roughly 6.4 mln USD. 

After the money cost estimation, we reanalyzed the results.  First, since the ML task we consider is essentially binary classification, we can simply flip the labels
of the analyzed data points. This way we're starting with the classifier
with 67\% accuracy, instead of 33\%. Moreover, since we fixed the
random seeds at the beginning of the experiment, we can simulate
the exact same untrained model we ran on the IBM machine. This way we could obtain  the data required for the comparison of the results obtained with simulator and the quantum device , thus allowing for a qualitative analysis of the results. In what follows, we present an analysis of our results, after the data relabeling.
Simulated results concern untrained model, the same one we used on the
real device, unless otherwise stated.

The binary classification accuracy we obtained was 67.44\% and 56.70\% for the
model ran on the simulator and the real device respectively. The accuracy
dropped by 10 percent points (or roughly 15\%). That's a significant drop,
especially when we consider that the model accuracy
isn't really impressive to begin with. It was, however, to be
expected. The quantum layer of our model consists of 3 layers of a
\texttt{SimplifiedTwoDesign} each built using several 2-qubit and
single qubit gates. Increasing the number of gates in the circuit raises the
computation time and thus influence of the decoherence. 
The decoherence brings the qubits' quantum state closer to
the maximally mixed state\footnote{While that's not always true, the noise model
used by IBM assumes that each gate is followed by (among others) a depolarizing
channel which introduces the described effect.}, hence
the results of the classification are also more random (accuracy closer to 50\%).
We'd also like to point out, that the same qualitative conclusions could be made
without the data relabeling. 

We then computed the values of the proposed explainability metrics, which we present in Table \ref{tab:metrics}. The idea was to check whether our expectations regarding the noise influence on the classification are reflected in the metrics. 


\bgroup
\def\arraystretch{1.5}
\setlength\tabcolsep{5pt}
\begin{table}[ht!]
	\caption{The value of accuracy and the explainability metrics for our model run on the noiseless simulator and on the \texttt{ibm\_brisbane} device. The worst value for sureness is 0.5 --- it says that the model was essentially guessing the outcomes. The closer confidence are to 1, the better. The values of imbalance is bounded by the number of shots and describes the tendency of a model towards a particular guess. It should be close to 0 for balanced data. See Sec. \ref{sec:metric} for more details. In the ideal scenario all the metrics should be equal (within statistical error) for both devices.}
	\label{tab:metrics}
	\centering
	\begin{tabular}{| c | c | c | c | c | c | c |} 
		\hline
		\textbf{Device} & \textbf{Accuracy} & \textbf{Sureness} & \textbf{Confidence} & $n_0$ & $n_1$ & \textbf{Imbalance} \\
		\hline
		Simulator & 0.674 & 0.118 & $0.523 \pm 0.064$ & 557 & 667 & -110  \\
		\hline
		\texttt{ibm\_brisbane} & 0.567 & 0.298 & $0.510 \pm 0.181$ &  192 & 839 & -647\\
		\hline
	\end{tabular}
\end{table}
\egroup

First, we looked at the Sureness. Our intuition was that due to the noise the model would drift toward random classifier, making the Sureness of the device-ran model closer to 0.5. However, that was not the case. This
result could hint at some kind of inertia-like or damping / relaxation effect
occurring on IBM devices. However, further research would be required to support this hypothesis.

We then looked at Confidence metric.  First we can see
that there's a drop in the model confidence. This is surely related to the
accuracy drop between the model ran on the simulator and the real
device. However, in the light of significant Sureness increase between
the same two models, it would seem that something more is going on.

To investigate this issue further, we looked upon the imbalance metric.
The results show that both models had a tendency to guess
class 1 better, but for a model ran on the real device, this tendency was
significantly higher. An important thing to note here is the fact that
we did the relabeling before the analysis. This means that class 1
predictions are made when the output of the model is low, that is the 
$\ket{1}$ state counts were low. Imbalance analysis led us to believe that we're
experiencing a significant influence of the relaxation-type \cite{Koch_2007} decoherence
on the real device. This hypothesis is further supported by the observation
that Qiskit devices' error model is constructed using thermal relaxation
error channels. However, further research in this area would be needed
to support this claim. We are, however, of firm believe, that running our model with trained weights would not lead to any new and interesting insights. Perhaps it would be interesting to revisit those experiments after device-oriented training becomes possible.

Our code and the data we gathered are available on Gitlab \cite{Code}
and Zenodo \cite{Zenodo}, respectively.

\section{Conclusions}

In this work we presented a case study of running a QML model on a quantum hardware. The model was meant to be trained specifically for the target device (device-oriented training) for pixel-level change detection on a pair of multispectral satellite images (binary classification). Such training was meant to limit the models error, by considering the device's topology and native gates set and increase the models quality by taking into account the noise model of the device. We also proposed new XAI techniques for QML that allowed us to explain the differences in the results from the simulation and the real hardware run. We also mentioned the software issues we encountered during the study.

The conclusion we draw from these experiments is that the available
software frameworks, i.e. PennyLane and Qiskit, do not fully support noisy
simulation of quantum machine learning models. They not only over-complicate
the process by hiding the physical-virtual qubit correspondence, but also
cannot support the simulations due to computational complexity. In case
of PennyLane, even the noise model itself is not complete (it lacks 
readout errors). To some limited extent, however, it seems possible to 
at least run the noisy forward pass for single samples.  With some work, that might allow implementation of a gradient free training of QNN \cite{Meta} with a noisy simulation. In any case, despite its highlighted advantages, the implementation of standard (gradient-based) device-oriented training seemed impossible. At least for our model and with software libraries we used.

These libraries are also subject to technical errors, which have to be
either corrected manually or submitted as an issue to their respective
developers. In both cases, the process is time-consuming and reduces 
the usability and confidence we can have in those libraries. This highlights
the need for a discussion on a status of the current QML software.

Our results also suggest that the QML models ran on real devices are
highly unreliable, inaccurate and costly. The cause of this behavior is undoubtedly the noise occurring during data processing on real devices. We believe that current quantum devices (at least the ones of similar class to the one we tested) are not ready for running complex ML models. Especially one trained in a noiseless environment. Nonetheless, in the advent of fault tolerant quantum devices, quantum computing (or quantum technologies) still remains a viable approach for addressing some of the ISE issues. One just has to remember that the application of those techniques might not always be as straightforward as it initially seems.

\section*{Acknowledgment}

Supported by ESA under the contract
No.~4000137375/22/NL/GLC/my. Supported by PMW programme under the
contract No. 5304/ESA/2023/0. The simulations were carried out based on the IBM
Quantum environment, access to which is financed within the framework of the
project “Development of digital competences in quantum engineering in
2024-2025” carried out by the Poznan Supercomputing and Networking Center.
PG and TR gratefully acknowledge the funding support by program “Excellence initiative—research university” for the AGH University in Krakow as well as the ARTIQ project: UMO-2021/01/2/ST6/00004 and ARTIQ/0004/2021.

\appendix

\section{Analyzed QNN layers}
\label{app:layers}

In this section, we present all the data encoding schemes and all the QNN layers analyzed during our experiments. We used two different kinds of data encoding: amplitude embedding and angle encoding. For the latter, we used two different sets of rotation gates: $RX$ and $RY$. We present the data encoding schemes we used in Fig. \ref{fig:enc}. Amplitude embedding is missing, from the figure because it is typically depicted as a single block. Possible implementation of amplitude embedding in the form of quantum circuit exploits the Mottonen State Preparation \cite{amplitude,Mottonen}.

\begin{figure*}[ht!]
	\centering
	\begin{subfigure}[t]{0.49\textwidth}
		\centering
		\begin{tikzpicture}[scale=1.3]
			\begin{yquant*}
				init {} q[4];
				box {RY} q[0-4];
			\end{yquant*}
		\end{tikzpicture}
	\end{subfigure}
	\begin{subfigure}[t]{0.49\textwidth}
				\centering
		\begin{tikzpicture}[scale=1.3]
			\begin{yquant*}
				init {} q[4];
				box {RX} q[0-4];
			\end{yquant*}
		\end{tikzpicture}
	\end{subfigure}
	\caption{Four qubit examples of angle encoding layers. Both RX and RY takes one parameter.}
	\label{fig:enc}
\end{figure*}

We used three different kinds of QNN layers: Strongly Entangling Layers, Simplified Two-Design and Bellman layers. We present the first two in Fig. \ref{fig:qnn_layers} and the Bellman layer in Fig. \ref{fig:bellman}.
\begin{figure*}[ht!]
	\centering
	\begin{subfigure}[t]{0.39\textwidth}
		\centering
	\begin{tikzpicture}[scale=1.3]
		\begin{yquant*}
			init {} q[7];
			box {Z} q[1] | q[0];
			box {Z} q[3] | q[2];
			box {Z} q[5] | q[4];
			box {Z} q[7] | q[6];
			box {RY} q[0-6];
			box {Z} q[2] | q[1];
			box {Z} q[4] | q[3];
			box {Z} q[6] | q[5];
			box {RY} q[1-7];
		\end{yquant*}
	\end{tikzpicture}
	\caption{Simplified two-design.}
	\end{subfigure}
	\begin{subfigure}[t]{0.59\textwidth}
		\centering
		\begin{tikzpicture}[scale=1.3]
			\begin{yquant*}
				init {} q[7];
				box {ROT} q[0-7];
				box {Z} q[1] | q[0];
				box {Z} q[2] | q[1];
				box {Z} q[3] | q[2];
				box {Z} q[4] | q[3];
				box {Z} q[5] | q[4];
				box {Z} q[6] | q[5];
				box {Z} q[7] | q[6];
			\end{yquant*}
		\end{tikzpicture}
		\caption{Strongly Entangling Layer ($r=1$).}
	\end{subfigure}
	\caption{8 qubit examples of QNN layers used in the experiment. $RY$ gate takes one parameter. $ROT$ takes 3 parameters and is a general rotation gate.}
	\label{fig:qnn_layers}
\end{figure*}

\begin{figure*}[ht!]
	\hspace*{-0.5cm}
	\begin{tikzpicture}[scale=1.3]
		\begin{yquant*}
			init {} q[7];
			H q[0];
			CNOT q[1] | q[0];
			CNOT q[2] | q[1];
			CNOT q[3] | q[2];
			CNOT q[4] | q[3];
			CNOT q[5] | q[4];
			CNOT q[6] | q[5];
			CNOT q[7] | q[6];
			box {RY} q;
			CNOT q[7] | q[6];
			CNOT q[6] | q[5];
			CNOT q[5] | q[4];
			CNOT q[4] | q[3];
			CNOT q[3] | q[2];
			CNOT q[2] | q[1];
			CNOT q[1] | q[0];
		\end{yquant*}
	\end{tikzpicture}
	\caption{A single Bellman layer. The $RY$ gates
		are parametrized by the QNN weights.}
	\label{fig:bellman}
\end{figure*}

\bibliographystyle{plain}
\bibliography{qml_case_study}

\end{document}